\documentstyle[aps]{revtex}
\begin{document}
\draft
\title{On Eigenvalue Decompactification in QCD$_{1+1} $ }
\author{Michael Engelhardt\thanks{email: hengel@wicc.weizmann.ac.il} }
\address{Department of Condensed Matter Physics \\
Weizmann Institute of Science \\ Rehovot 76100, Israel }
\date{}
\maketitle

\begin{abstract}
The question whether it is necessary to decompactify the gauge
eigenvalue degrees of freedom in QCD$_{1+1} $ is addressed. A
careful consideration of the dynamics governing these degrees
of freedom leads to the conclusion that eigenvalue decompactification
is not necessary due to the curvature on the space of eigenvalues.
\end{abstract}
\vspace{1cm}

\pacs{Keywords: QCD in one space dimension, Quantization on curved
manifolds.}

Recently, concerns have been raised regarding a possible
requirement of ``eigenvalue decompactification'' \cite{wad1}
\cite{wad2} in QCD in one space and one time dimension (QCD$_{1+1} $).
The emergence of such a requirement would e.g. in the limit of a
large number of colors cast serious doubts upon the validity of
the usual large $N$ counting arguments \cite{wad2}. The purpose of this
note is to reexamine carefully the dynamics governing the abovementioned
eigenvalues; this will lead to the conclusion that eigenvalue
decompactification is not necessary.

In order to familiarize the reader with the issue, a short review
of the methodology of \cite{wad1},\cite{wad2} is in order. \cite{wad1},
\cite{wad2}, along with many other modern treatments of 1+1-dimensional
gauge theories, choose the spatial coordinate to be compactified to
a ring with tunable circumference $L$ in order to obtain good control
of the infrared properties of the model. On such a manifold, the gauge
degrees of freedom cannot be completely gauged away; one must keep e.g.
the zero momentum mode of the spatial component of the gauge field
$A_1 $. One may further constrain this mode to be diagonal, which
leaves one with $N-1$ quantum mechanical gauge degrees of freedom in
an $SU(N)$ theory \cite{lenz}; these are the eigenvalues alluded to
above (one of the eigenvalues, say in the following the $N$th, is
constrained due to the tracelessness of the $SU(N) $ generators).
In addition, one has the original fermion fields as physical
variables. The Hamiltonian can be written as
\begin{equation}
H=H_{KIN} + H_F + H_{COUL}
\end{equation}
where $H_{KIN} $ is the kinetic energy of the eigenvalues, $H_F $
is the usual kinetic piece for the fermions, minimally coupled
to the eigenvalues, and $H_{COUL} $ is the Coulomb interaction,
which also depends on the eigenvalues \cite{lenz}.

When quantizing the theory, one must take care to properly account for
the curvature on the space of the eigenvalues. The measure in the
scalar product of the Hilbert space of eigenvalue wave functions is
the eigenvalue part of the Haar measure of $SU(N)$,
\begin{equation}
J=\prod_{i<j} \sin^{2} \left( \frac{\lambda_{i} -\lambda_{j} }{2} \right)
\label{jac}
\end{equation}
where the $\lambda_{i} $ are the $N$ eigenvalues of $A_1 $ \cite{doug}
\cite{mina}. This is easy to understand: Since in the gauge sector, only
the momentum zero mode of the vector field is relevant, QCD$_{1+1} $ can
inherently be thought of as a one-link Hamiltonian lattice gauge theory
(with periodic boundary conditions) \cite{kogsus}. The gauge degree
of freedom parametrizes the link variable $U$ in the canonical
fashion\footnote{The gauge degree of freedom $A_1 $ used here is
already rescaled by a factor $gL$ with respect to the $A_1 $ one usually
writes in the QCD Lagrangian.},
\begin{equation}
U=e^{iA_1 }
\end{equation}
i.e. it parametrizes the $SU(N) $ rotation of the color frame of
reference if one moves once around the spatial circle.
Note that the fermions, on the other hand, may be allowed to move
continuously along the link since, in one dimension, there is no
path ordering ambiguity; a Wilson line is completely specified
by its endpoints (up to the number of times it winds around the spatial
circle). Specifically, if one moves a distance $x$ in space, the
color frame of reference is rotated by\footnote{A nice example of
the intricate interplay between the gauge phases picked up by the
fermions and the eigenvalue degrees of freedom is given in
\cite{selbst}.}
\begin{equation}
U_x = e^{iA_1 x/L}
\end{equation}
Thus it becomes clear that the appropriate measure for the gauge
degree of freedom is the Haar measure corresponding to the gauge group.
Choosing a gauge in which $A_1 $ is diagonal quantum mechanically
corresponds to restricting the Hilbert space to wave functions
independent of the angular parts of $A_1 $, leaving only the
eigenvalue part (\ref{jac}) as the relevant measure on the space
of physical states.

Corresponding to the measure (\ref{jac}), the Laplacian entering the
kinetic energy $H_{KIN} $ acquires a Jacobian \cite{doug},
\begin{equation}
\Delta = \sum_{i=1}^{N-1}
\frac{1}{J} \frac{\partial }{\partial \lambda_{i} }
J \frac{\partial }{\partial \lambda_{i} }
\end{equation}

Having established the form of the operators, one can now focus on
the range of values the $\lambda_{i} $ may take. The pure gauge
theory is completely periodic in the eigenvalues, with period $2\pi $;
thus, one may think of this theory as a system of $N-1$ particles
living on a circle.
On the other hand, when fermions are included, the theory is only
symmetric if, simultaneously to shifting an eigenvalue by $2\pi $,
one gives the appropriate phase to the fermions. Thus, in general,
one will have to allow for an infinite range of values for the
$\lambda_{i} $. This is what happens e.g. in the Schwinger model
\cite{mant}, where one must superimpose vacua related by large
gauge transformations to form the well-known $\theta $-vacua.

Now, however, one must take into account the Jacobian on the
Hilbert space of eigenvalue wave functions discussed
above\footnote{The importance of such Jacobians has also been
emphasized e.g. in \cite{le1}, \cite{le2}.}.
This Jacobian prevents quantum mechanical propagation past the points
where it is zero. One can see this clearly e.g. by constructing the
probability current for the $\lambda_{i} $ in the usual way, by demanding
conservation of probability \cite{merz}. The probability current is
proportional to $J$, and thus vanishes whenever two of the $N$ eigenvalues
meet modulo $2\pi $. Since furthermore, their center of mass is fixed,
they cannot propagate out of a certain compact fundamental domain. Note
that this argument remains valid when taking into account the
fermions, as long as the Hamiltonian remains local in the space of
eigenvalues and does not contain attractive singular potentials which
could induce the eigenvalues to fall towards each other. Both of these
conditions are fulfilled. It should be mentioned that the Coulomb interaction
indeed contains quadratic singularities at the zeros of the Jacobian
\cite{wad1} \cite{selbst}; they however always have a positive coefficient,
essentially the square of the fermionic color charge. Thus, except for
fermionic color charge zero states, the effect of the Jacobian is even
reinforced.

The different fundamental domains are consequently dynamically decoupled,
there is no possibility of quantum mechanical interference like in
the Schwinger model, which contains no Jacobian\footnote{Correspondingly,
whereas the Schwinger model contains a nontrivial continuous $\theta $-angle,
which can be interpreted as the Bloch momentum of propagation around
the $A_1 $ circle, in the $SU(N) $ theory the Bloch momentum disappears
due to the Jacobian barriers. This was first pointed out in \cite{wit}.
Note that for quarks in the adjoint representation, there are actually
additional symmetries \underline{within} the fundamental domain which
allow the construction of $N$ different $\theta $-vacua. However, one
can never obtain a continuous band of $\theta $-vacua \cite{wit} as would
be produced if the shifts of the eigenvalues by $2\pi $ were nontrivial.}.
One can give a complete basis of eigenfunctions of the full
Hamiltonian in terms of wave functionals which vanish on all but
a single fundamental domain in the eigenvalues. Furthermore, there is
no observable in the theory which connects different fundamental
domains\footnote{In particular, the generator of translations of
the variable $\lambda_{i} $, namely $i\partial / \partial \lambda_{i} $,
is not hermitian in the presence of the measure (\ref{jac}); therefore,
the quantum mechanical generator of the corresponding residual gauge
transformations, which
in the case of the Schwinger model has the $\theta $-angle as its
eigenvalue, is not an observable in the present case.}.
This means that, though one may define the eigenvalues to have an
infinite range to begin with, one is in fact adding up trivial copies
of the theory defined on one fundamental domain. One may just as well
restrict to one of these domains only, factoring out the (infinite)
number of domains; invariance under the corresponding residual gauge
transformations simply ceases to be an issue.

In concluding, three further comments are in order: First, the
considerations above were for an $SU(N) $ theory, which fixes the
center of mass of the eigenvalues. If one considers a $U(N) $ theory,
the center of mass is a further degree of freedom. This degree of
freedom, and it alone, is not constrained to a compact domain by
the Jacobian. Thus, one obtains the same $U(1) $ anomaly as in the
Schwinger model. However, for all the relative coordinates, i.e.
the eigenvalues with the center of mass subtracted off, the same
discussion as above applies. The subtleties of the $N\rightarrow \infty $
limit connected with the question whether one includes the $U(1) $
part in the gauge group or not have been discussed e.g. by
D.Stoll \cite{thesis}.

Secondly, the Jacobian (\ref{jac}) is often treated in the following
way: By going to ``radial'' wave functions
\begin{equation}
\psi (\vec{\lambda } ) = \prod_{i<j} \sin \left(
\frac{\lambda_{i} -\lambda_{j} }{2} \right) \phi (\vec{\lambda } )
\label{rad1}
\end{equation}
one may eliminate the Jacobian from the scalar product and the Laplacian
(up to an irrelevant overall constant in the energy). The only, however
important, relic of the curvature on the space of eigenvalues lies in
the fact that $\psi (\vec{\lambda } ) $ now must vanish at certain points.
Then the $N$ eigenvalues are interpreted as fermionic degrees of
freedom\footnote{Note that this contains all the information about the
nodes of the wave function only as long as one has the eigenvalues
living on a circle, since the Jacobian vanishes always when two
eigenvalues coincide modulo $2\pi $. If one allows an infinite range
for the eigenvalues, one must additionally remember to impose nodes
when two eigenvalues match up to a multiple of $2\pi $. This crucial point
was neglected in \cite{wad2}.} \cite{mina} (with fixed center of
mass in the $SU(N) $ case), since interchange leads to a minus sign in
(\ref{rad1})\footnote{Note that $\phi (\vec{\lambda } ) $ can be argued
to be symmetric in the $\lambda_{i} $, reflecting the fact that it
should not matter in which order one puts the eigenvalues when
diagonalizing $A_1 $.}. There is nothing wrong with such an interpretation;
it is one particular way of superimposing wave functions in different
fundamental domains. One should however not be misled into thinking
that this is the only legitimate way. Equally possible is e.g.
\begin{equation}
\psi (\vec{\lambda } ) = \sqrt{J} \, \, \phi (\vec{\lambda } )
\label{rad2}
\end{equation}
Note that the additional $\delta $-functions one picks up when acting
with the kinetic energy on (\ref{rad2}) as opposed to (\ref{rad1})
always coincide with zeros of the wave function. Thus, whether one
chooses (\ref{rad1}), (\ref{rad2}), or restricts to a fundamental
domain, one will always recover the same unique answer for physical
quantities.

Finally, it should be noted that the conclusions reached above can also
be made plausible in the path integral formalism. Consider for
simplicity $SU(2) $, where there is only one independent eigenvalue
$\lambda $. The points $\lambda = n\pi $ with integer $n$ where the
Jacobian vanishes correspond to poles of the four-dimensional sphere
to which $SU(2) $ is isomorphic. Thus, heuristically, a path which
leads past a zero of the Jacobian would have to lead exactly over a
pole of the sphere. The set of such paths has measure zero.
Formally, the Jacobian enters as integration measure of the
$\lambda_{i} $ at each time slice. Therefore, any path which leads over
a zero of the Jacobian gives a vanishing contribution to the
propagator.

\subsection*{Acknowledgements}
Discussions with F.Lenz, A.Schwimmer, and especially B.Schreiber
are gratefully acknowledged.
This work was supported by a MINERVA fellowship.

\end{document}